\documentclass[conference]{IEEEtran}
\IEEEoverridecommandlockouts

\usepackage{cite}
\usepackage{amsmath,amssymb,amsfonts}
\usepackage{algorithmic}
\usepackage{graphicx}
\usepackage{textcomp}
\usepackage{xcolor}
\usepackage{cite}
\usepackage{hyperref} 
\usepackage{url}
\usepackage{float}
\usepackage{booktabs}
\usepackage{makecell}
\usepackage{pifont}
\def\BibTeX{{\rm B\kern-.05em{\sc i\kern-.025em b}\kern-.08em
    T\kern-.1667em\lower.7ex\hbox{E}\kern-.125emX}}

\begin{document}

\title{UavNetSim-v1: A Python-based Simulation Platform for UAV Communication Networks}
\author{
	\IEEEauthorblockN{
		Zihao Zhou\IEEEauthorrefmark{1},
		Zipeng Dai\IEEEauthorrefmark{2},
		Linyi Huang\IEEEauthorrefmark{3},
		Cui Yang\IEEEauthorrefmark{1},
		Youjun Xiang\IEEEauthorrefmark{1},
		Jie Tang\IEEEauthorrefmark{1},
		Kai-kit Wong\IEEEauthorrefmark{4}\IEEEauthorrefmark{5}
	}
	\IEEEauthorblockA{\IEEEauthorrefmark{1}School of Electronic and Information Engineering, South China University of Technology, Guangzhou, China}
	\IEEEauthorblockA{\IEEEauthorrefmark{2}Department of Computer Science and Technology, Beijing Institute of Technology, Beijing, China}
	\IEEEauthorblockA{\IEEEauthorrefmark{3}Thrust of Robotics and Autonomous Systems, \\
		The Hong Kong University of Science and Technology (Guangzhou), Guangzhou, China}
	\IEEEauthorblockA{\IEEEauthorrefmark{4}Department of Electrical and Electronic Engineering, University College London, London, United Kingdom}
	\IEEEauthorblockA{\IEEEauthorrefmark{5}Yonsei Frontier Lab, Yonsei University, Seoul, Korea}
	\IEEEauthorblockA{Corresponding author: Jie Tang, eejtang@scut.edu.cn}
}

\maketitle

\begin{abstract}
In unmanned aerial vehicle (UAV) networks, communication protocols and algorithms are essential for cooperation and collaboration between UAVs. Simulation provides a cost-effective solution for prototyping, debugging, and analyzing protocols and algorithms, avoiding the prohibitive expenses of field experiments. In this paper, we present ``UavNetSim-v1'', an open-source Python-based simulation platform designed for rapid development, testing, and evaluating the protocols and algorithms in UAV networks. ``UavNetSim-v1'' provides most of the functionalities developers may need, including routing/medium access control (MAC) protocols, topology control algorithms and mobility/energy models, while maintaining ease of use. Furthermore, the platform supports comprehensive performance evaluation and features an interactive visualization interface for in-depth algorithm analysis. In short, ``UavNetSim-v1'' lends itself to both rapid prototyping and educational purposes, and can serve as a lightweight yet powerful alternative to mature network simulators for UAV communication research.
\end{abstract}


\section{Introduction}
Unmanned aerial vehicle (UAV) networks have recently attracted widespread attention from both academia and industry, playing an increasingly vital role in diverse applications such as post-disaster rescue\cite{N.Zhao-2019}, cooperative reconnaissance\cite{X.Li-2024} and traffic monitoring \cite{B.Yang-2023}. The execution of complex tasks relies on communication and collaboration among UAVs, which depends on the design of appropriate network protocols (e.g., routing protocols, medium access control (MAC) protocols) and control algorithms (e.g., motion control algorithms, resource allocation algorithms) \cite{S.Javed-2024}.

\begin{figure}[t]
	\centering
	\includegraphics[width=3.4in]{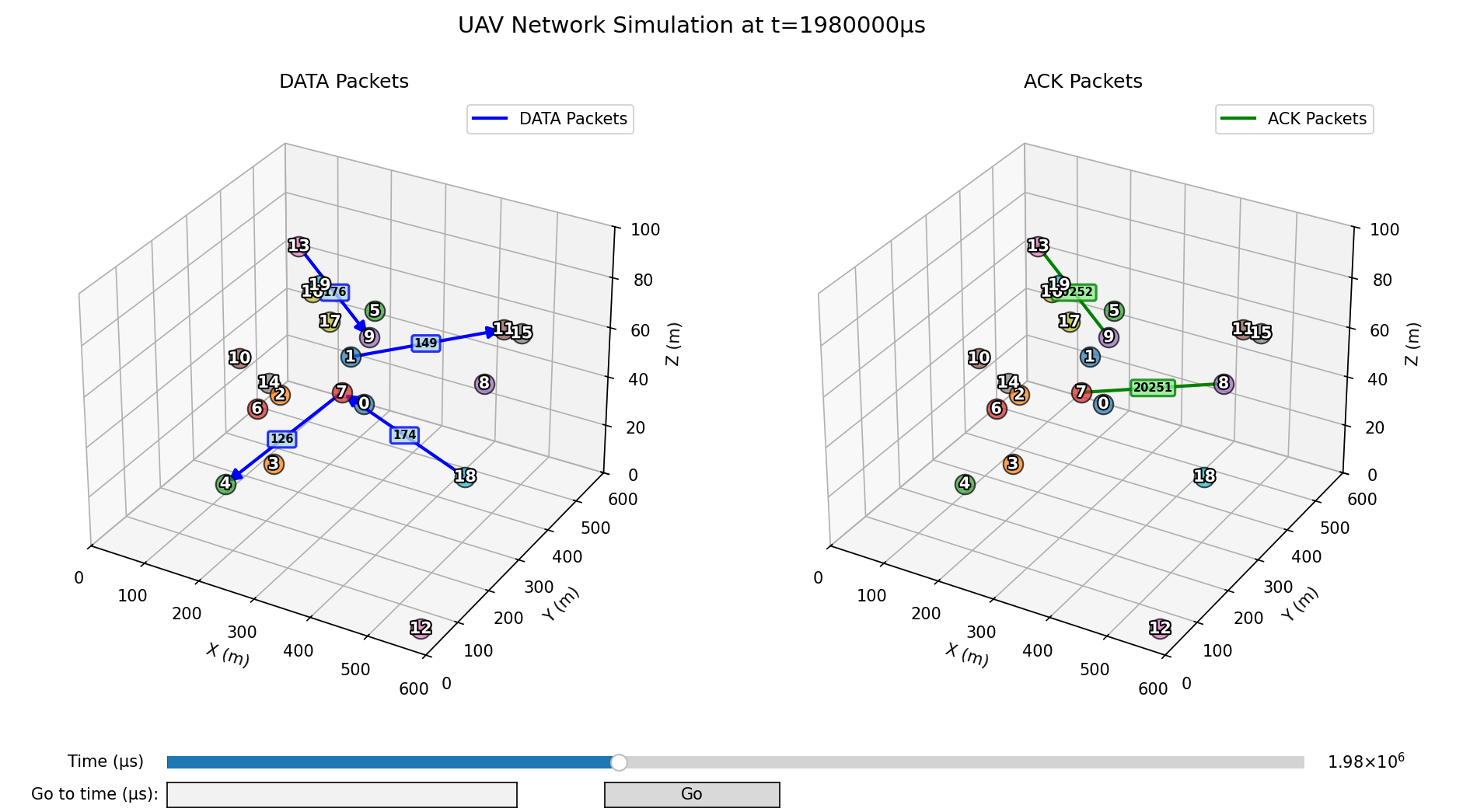}
	\caption{Interactive visual analysis window of ``UavNetSim-v1''.}
	\label{figure_1}
\end{figure}

\begin{figure*}[t]
	\centering
	\includegraphics[width=6.0in]{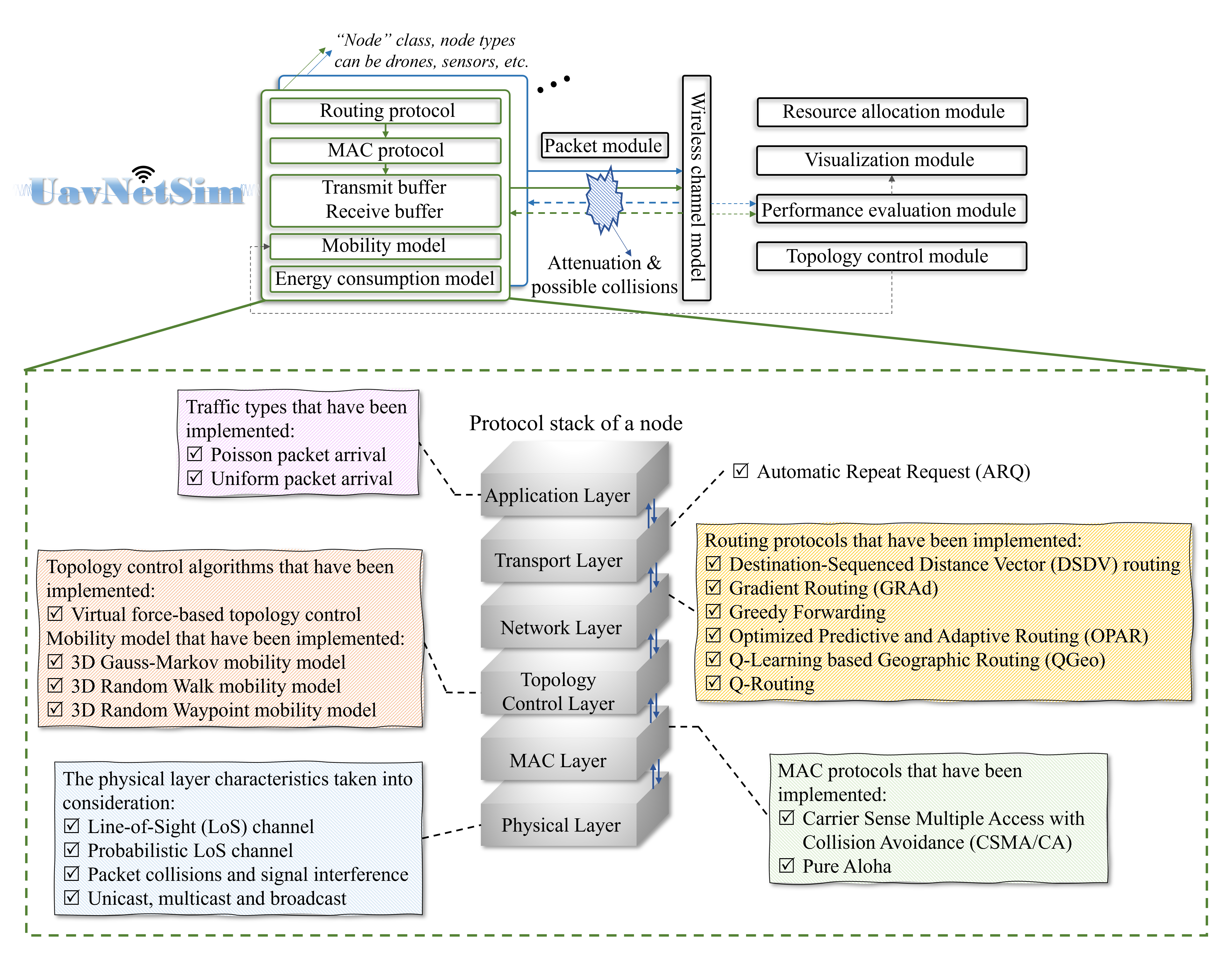}
	\caption{Architecture of ``UavNetSim-v1''.}
	\label{figure_2}
\end{figure*}

In prevailing research methodologies, the prototypes of protocols and algorithms for UAV networks typically requires simulation-based validation, as real-world experimentation is often cost-prohibitive. Beyond cost considerations, compared with real-world implementation, simulation platform is not only more convenient for prototyping, tuning, debugging and performance analysis of protocols and algorithms, but also ensures reproducibility under different conditions. Several open-source frameworks have been developed for simulating UAV \cite{J.Meyer-2012, O.Michel-2004, E.Rohmer-2013, S.Shah-2018}, however, the majority of these existing simulators focus on single-UAV fidelity and cannot support real-time simulation for large-scale multi-UAV systems, making them unsuitable for UAV communication network studies. On the other hand, for wireless communication networks, there are also many available simulators like GloMoSim \cite{X.Zeng-1998}, NS-2 \cite{T.Issariyakul-2009}, NS-3, OMNeT++, etc. Nevertheless, these simulators strive to achieve a perfect representation of all network layers, resulting in significant conceptual complexity and steep learning curve for beginners. Given the intricacy of such simulators, fully comprehending their entire functionality is nearly impossible, which may lead to improper use and, consequently, erroneous or misleading simulation results. Therefore, when used for rapid prototyping or educational purposes, the complexity of network simulators mentioned above would become a significant drawback. The time and effort required to master a full-featured simulator often outweigh the benefits.

Against these backgrounds, in this paper, we present a new open-source simulation platform ``UavNetSim-v1''\footnote{Our ``UavNetSim-v1'' simulation platform is now available on Github: \url{https://github.com/Zihao-Felix-Zhou/UavNetSim-v1}} for UAV communication networks. ``UavNetSim-v1'' is developed entirely in Python, designed primarily for rapid prototyping and educational purposes. This platform enables design and verify various protocols and algorithms such as routing protocols, MAC protocols and topology control in UAV networks while providing comprehensive performance evaluation and interactive visual analysis. Additionally, a modular programming approach is adopted in ``UavNetSim-v1'', making it as easy as possible for beginners to understand while allowing experienced users to flexibly extend the platform as needed.

The remainder of the paper is organized as follows: Section II summarizes the design goals and key features of ``UavNetSim-v1'', Section III describes the architecture of the simulation platform and gives an overview of different modules. Section IV demonstrates how ``UavNetSim-v1'' can be further extended. Finally, conclusions of this paper is given in Section V.

\section{Design Goals and Key Features}
As mentioned in Section I, the core objective of designing the ``UavNetSim-v1'' platform is to make the simulator easier to use and extend while maintaining high fidelity as much as possible. The key features of our platform can be summarized as follows: 
\begin{itemize}
	\item \textbf{Open-source}: The ``UavNetSim-v1'' project adopts community-oriented open
	source development practices.
	\item \textbf{Python-based}: ``UavNetSim-v1'' is entirely written in Python, the very fast learning curve of Python makes this platform easier to get started with.
	\item \textbf{Simple and flexible}: ``UavNetSim-v1'' is easy to use and extend to different application scenarios (e.g., flying ad-hoc networks (FANETs) \cite{I.Bekmezci-2013}, space-air-ground integrated network \cite{J.Liu-2018}).
	\item \textbf{Visualization-enabled}: As shown in fig. \ref{figure_1}, ``UavNetSim-v1'' can provide the visualization of UAV flight trajectory and packet forwarding path, which facilitates intuitive analysis of the behavior of algorithms and protocols.
	\item \textbf{AI-supported}: ``UavNetSim-v1'' enables design and verification of artificial intelligence (AI)-based protocols or algorithms (e.g., reinforcement learning (RL)-aided routing protocols \cite{J.Boyan-1994, W.Jung-2017}).
	\item \textbf{Performance analysis-enabled}: Support different metrics for performance evaluation (e.g., packet delivery ratio (PDR), end-to-end (E2E) delay, throughput, overhead and routing hop count).
\end{itemize}

\begin{figure*}[t]
	\centering
	\includegraphics[width=6.0in]{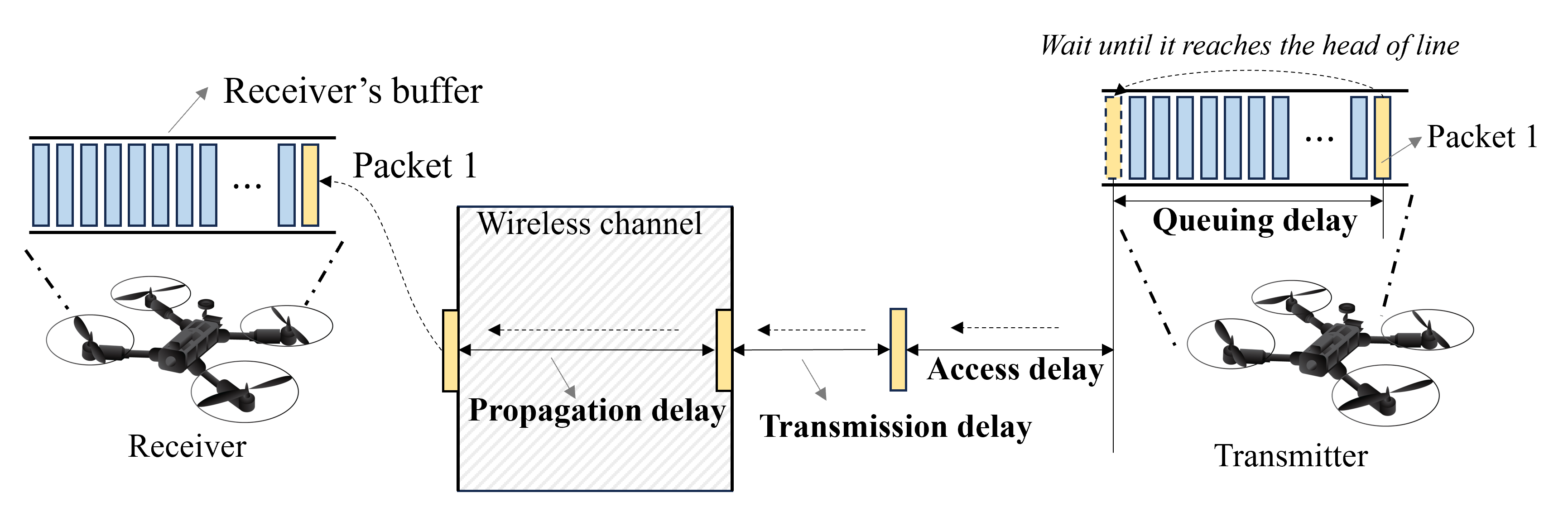}
	\caption{The delay composition of a successful single-hop packet transmission in ``UavNetSim-v1''.}
	\label{figure_3}
\end{figure*}

\section{Architecture and Modules Overview}
This section provides an overview of the architecture of ``UavNetSim-v1'' and some of the important modules and packages. ``UavNetSim-v1'' is developed based on SimPy \footnote{\url{https://simpy.readthedocs.io/en/latest/}}, a process-based discrete-event simulation framework based on standard Python \cite{D.Zinoviev-2024}. The core elements of SimPy's architecture include the Environment, Events \footnote{For example, the timeout events can be used to simulate the delays.}, and user-defined Process functions \footnote{The process functions represent the behavioral models intended for simulation}. Simpy implements the concept of Python generator function which suspends process when an event is yielded and resumes only when interrupted or triggered externally. The SimPy engine operates as an asynchronous scheduler, coordinating the execution of process functions based on events and schedules specified. Such a feature makes SimPy well suited for asynchronous networking and multi-agent communication systems.

In ``UavNetSim-v1'', ``Drone'' is an important class, which is defined in {\ttfamily entities/drone.py}. Our platform allows spreading a collection of UAVs on the three-dimensional (3D) area, whenever a UAV is added to the simulation environment, a ``Drone'' class is instantiated. As shown in fig. \ref{figure_2}, each UAV needs to install the protocol stack, including routing protocol, MAC protocol, mobility model, energy consumption model, etc. Since we are not trying to include all the characteristics of the application layer, so ``{\ttfamily generate\_data\_packet()}'' is served as a key process function of ``Drone'' class to simulate the distribution of data packet arrivals. 

When ``{\ttfamily generate\_data\_packet()}'' is executed, a ``DataPacket'' class will be instantiated. ``DataPacket'' class inherits from ``Packet'' base class, they are both defined in {\ttfamily entities/packet.py}. Then, when the data packet reaches the head-of-line (HOL) of the UAV, the UAV first decides the appropriate next hop according to the routing protocol. Next, it decides when to transmit the packet over the wireless channel according to the MAC protocol. Packets experience attenuation as well as possible collisions as they travel over the wireless channel. Note that ``UavNetSim-v1'' does not incorporate all physical transmission characteristics (e.g., encoding, modulation) to avoid undue complexity. When a packet successfully arrives at its destination, relevant performance metrics (see Section III.A) are recorded and an interactive visual analysis is provided at the end of the simulation.

Next, we will introduce some of the important modules or packages. For the accuracy of description, we use ``module'' to represent a single Python file that contains collections of functions and variables and with the extension ``.py''. ``Package'' is used to represent a directory having collections of modules.

\subsection{Performance Evaluation Module}
Before presenting the other functional packages, we briefly introduce the performance evaluation module in our platform. ``UavNetSim-v1'' supports the performance analysis of the UAV communication network, and the performance evaluation module is located in {\ttfamily simulator/metrics.py}. In the current version, five metrics are considered:
\begin{itemize}
	\item \textbf{Packet delivery ratio (PDR)}: the ratio of the data packets successfully received at the destinations to those generated at the sources.
	\item \textbf{Average end-to-end (E2E) delay}: the average time required for a successful data packet transmission between the source and destination, which includes all possible delays caused by queuing, contending for the channel, re-transmissions, transmission and propagation. In our platform, the composition of the delay of a successful single-hop packet transmission is shown in fig. \ref{figure_3}.
	\item \textbf{Average throughput}: it refers to the total amount of successfully received data at the destinations per unit time, it is measured in Kbps.
	\item \textbf{Routing load}: it is the number of control packets transmitted per data packet delivered at the destination.
	\item \textbf{Hop count}: it is the number of intermediate nodes (UAVs) a data packet traverses from the source to the destination in a network.
\end{itemize}

\subsection{Routing Protocol Package}
The routing protocol-related package in ``UavNetSim-v1'' is located in the {\ttfamily routing} folder, which in turn contains several sub-packages representing different routing protocols. In the current version, ``UavNetSim-v1'' supports destination-sequenced distance vector routing (DSDV) \cite{C.Perkins-1994}, greedy forwarding \cite{M.Khaledi-2018}, Gradient routing (GRAd) \cite{R.Poor-2000}, optimized predictive and adaptive routing (OPAR) \cite{M.Gharib-2021}, Q-routing \cite{J.Boyan-1994} and Q-learning-based geographic routing (QGeo) \cite{W.Jung-2017}.

For each class of routing protocol, two important member functions must be implemented:
{\ttfamily next\_hop\_selection()} and {\ttfamily packet\_reception()}. The former is to select the appropriate next hop node to forward the data packet according to a specific routing policy. The latter takes different actions according to the received packet type, such as updating the neighbor table, further forwarding, etc.

\begin{figure}[h]
	\centering
	\includegraphics[width=3.0in]{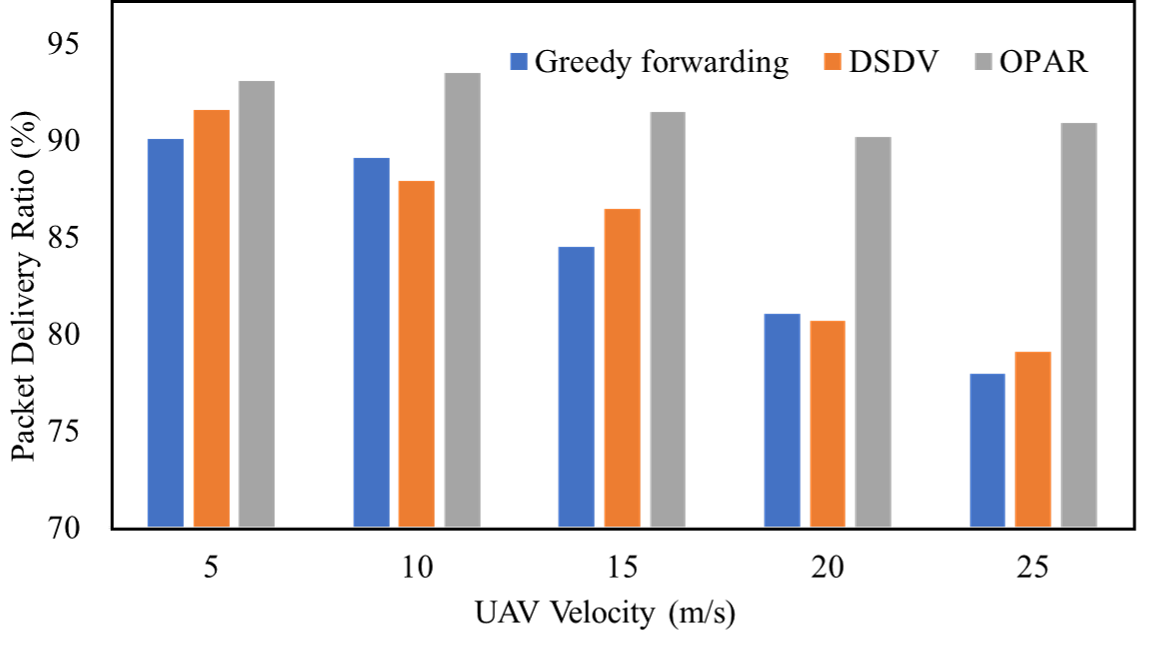}
	\caption{PDR for varying UAV velocities.}
	\label{figure_4}
\end{figure}

\begin{figure}[h]
	\centering
	\includegraphics[width=3.0in]{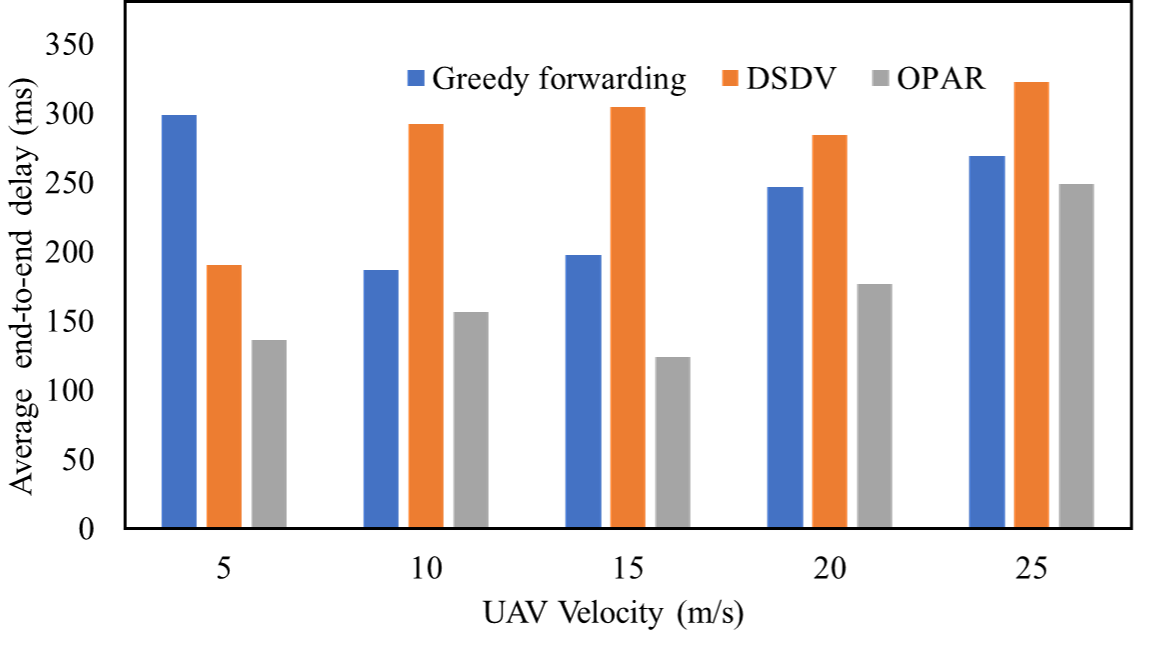}
	\caption{Average E2E delay for varying UAV velocities.}
	\label{figure_5}
\end{figure}

As an embodiment of a FANET, in fig. \ref{figure_4} and \ref{figure_5}, we compare the PDR and average E2E delay performance of three routing protocols implemented in ``UavNetSim-v1'': greedy forwarding, DSDV and OPAR, under different velocities of UAVs. The detailed parameters utilized in our simulation are compiled in Table I. Initially, 15 UAVs are randomly deployed in a 3-D space of 600m $\times$ 600m $\times$ 100m. The velocity of UAVs is 5-25 m/s. Each UAV can be served as source, relay or destination, and the source node generates data packets according to a Poisson process with parameter $\lambda$=5 packets/s. The UAV moves according to 3D Gauss Markov mobility model mentioned in Section III.D.

Fig. \ref{figure_4} demonstrates the PDR performance of the compared routing protocols for different UAV velocities. Due to its global observation of the network, OPAR outperforms both greedy forwarding and DSDV in terms of PDR. Moreover, it is also observed that the PDR performance of all protocols decreases to varying degrees as the UAV velocity increases. This is because higher UAV velocities lead to more rapid changes in network topology, resulting in frequent communication link disconnections, which in turn adversely affect routing decisions. Considering average E2E delay in fig. \ref{figure_5}, when the UAV velocity is 5 m/s, greedy forwarding has the highest delay (298.613 ms). As the velocity increases to 25 m/s, DSDV exhibits the highest delay (322.333 ms). On the other hand, OPAR can achieve the lowest average E2E delay at different UAV velocities.

\begin{table}[t]
	\centering
	\caption{EXPERIMENT PARAMETERS}
	\label{table1}
	\begin{tabular}{|c|c|}
		\hline
		Parameters & Settings \\
		\hline
		Map size & 600m $\times$ 600m $\times$ 100m \\
		Velocity of UAV & 5-25m/s \\
		UAV transmit power & 0.1W \\
		Antenna & Omnidirectional antenna \\
		Noise power & 4$\times 10^{-11}$ W \\
		SINR threshold & 6dB \\
		MAC protocol & CSMA/CA \\
		Carrier frequency & 2.4 GHz \\
		Bit rate & 2 Mbps \\
		Bandwidth & 22 MHz \\
		Time slot duration & 20 $\mu$s \\
		Short inter-frame space (SIFS) & 10 $\mu$s \\
		Distributed inter-frame space (DIFS) & 50 $\mu$s \\
		Initial contention window size & 31 \\
		Re-transmission limit & 5 \\
		IP header length & 20 bytes \\
		MAC header length & 14 bytes \\
		PHY header length & 24 bytes \\
		Data packet payload length & 1024 bytes \\
		ACK packet length & 30 bytes \\
		\hline
	\end{tabular}
\end{table}

\subsection{MAC Protocol Package}
The MAC protocol-related package in ``UavNetSim-v1'' is located in the {\ttfamily mac} folder, which in turns contains several sub-packages representing different MAC protocols. Carrier sense multiple access with collision avoidance (CSMA/CA) and pure ALOHA have been implemented in the current version. 

Similarly, for each MAC protocol, two member functions should be implemented: {\ttfamily mac\_send()} and {\ttfamily wait\_ack()}. The main role of {\ttfamily mac\_send()} is to determine the time at which a packet should be transmitted based on certain rules. The function {\ttfamily wait\_ack()} is responsible for waiting for an acknowledgement (ACK) packet from the next-hop UAV after transmitting a data packet \footnote{In scenarios where an ACK mechanism is not required, e.g., when using GRAd as the routing protocol, the {\ttfamily wait\_ack()} function does not need to be invoked.}. If no ACK packet is received within a specified time threshold, the transmission is considered a failure, and a re-transmission is initiated. If the maximum re-transmission limit is exceeded, the packet is dropped.

\subsection{Mobility Package}
\begin{figure}[h]
	\centering
	\includegraphics[width=3.5in]{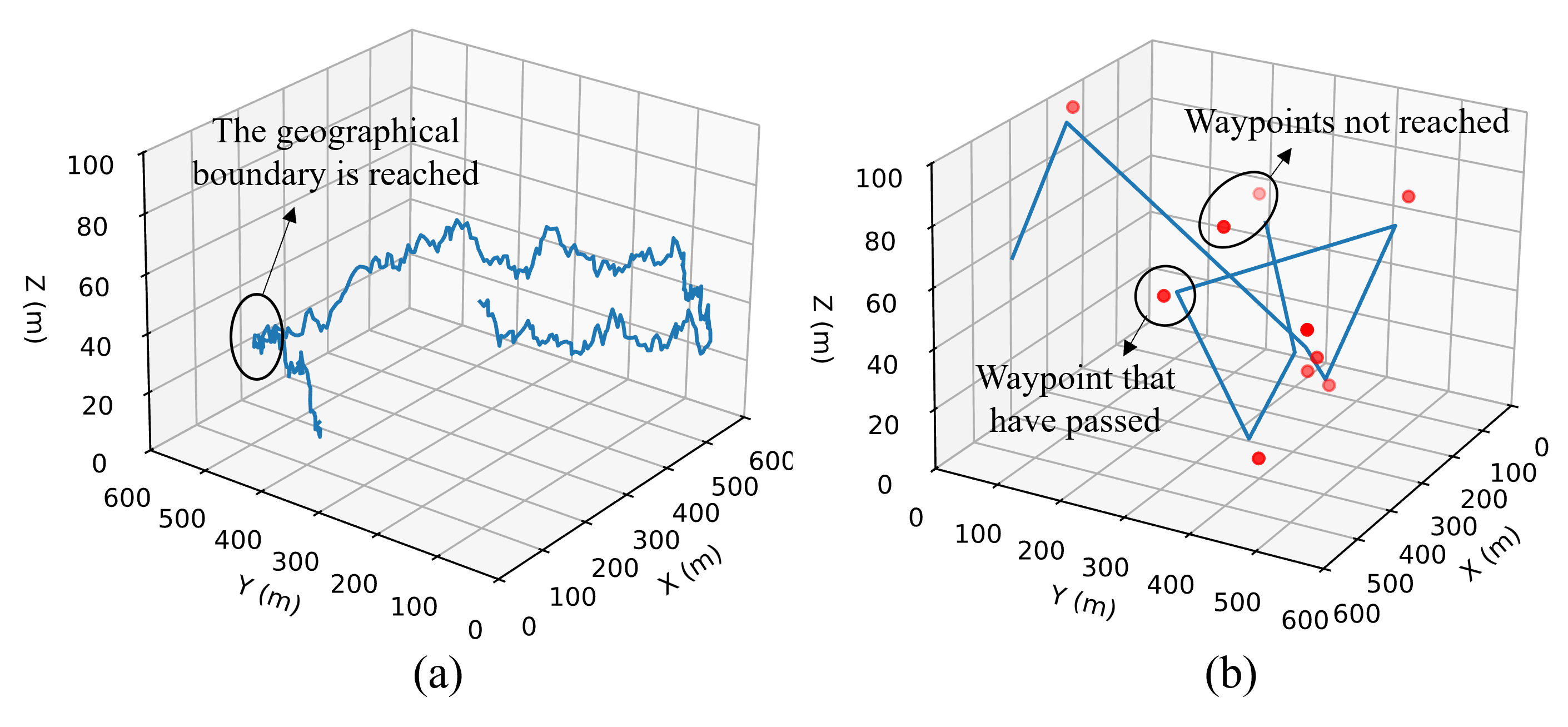}
	\caption{Traveling pattern of a UAV under (a) 3D Gauss Markov mobility model and (b) 3D random waypoint mobility model.}
	\label{figure_6}
\end{figure}

In ``UavNetSim-v1'', 3D Cartesian coordinate system is adopted, and the coordinates of UAV at time $t$ is $[x(t), y(t), z(t)]\in\mathbb{R}^{3\times 1}$. The folder {\ttfamily mobility} includes several different mobility modules: 3D Gauss-Markov mobility model, 3D random walk mobility model and 3D random waypoint mobility model. Each class of mobility model should have a member function named ``{\ttfamily mobility\_update()}'', which is used to update the locations, speed, direction and pitch of UAV node periodically. Fig. \ref{figure_6} demonstrates the movements of a UAV under 3D Gauss Markov mobility model and 3D random waypoint mobility model, where the red points in fig. \ref{figure_6}(b) represent the waypoints.

\begin{figure*}[t]
	\centering
	\includegraphics[width=7.0in]{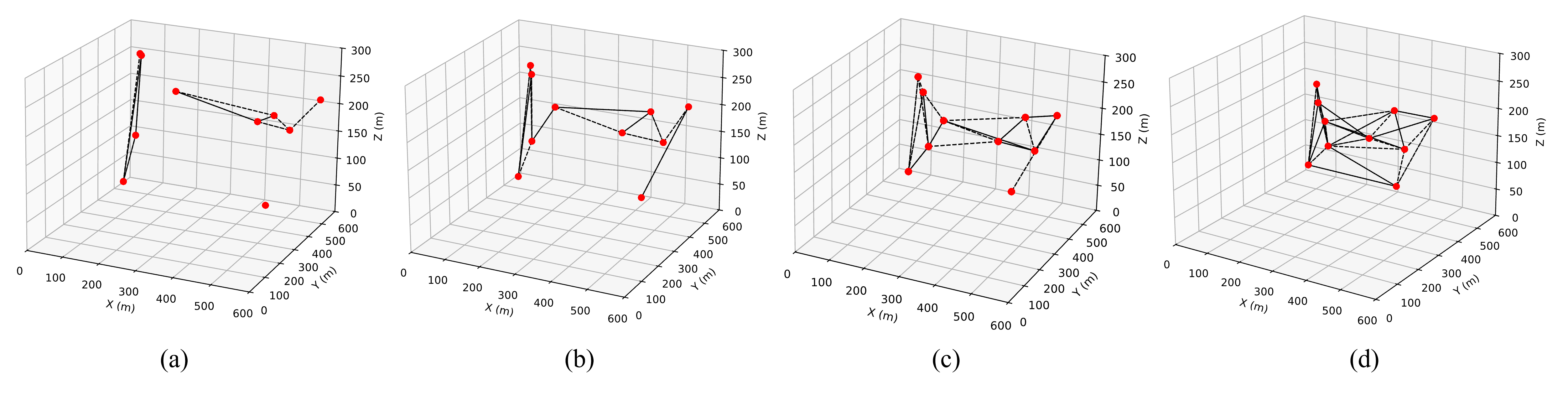}
	\caption{Dynamics of the movement process using virtual force-based motion control algorithm at (a) $t$=0s, (b) $t$=20s, (c) $t$=40s and (d) $t$=60s.}
	\label{figure_7}
\end{figure*}

\subsection{Topology Control Package}
The mobility models presented in Section III.D are mainly random mobility models, that is, the individual UAV nodes move randomly and independently among themselves. However, in real-world application scenarios, UAVs often need to realize cooperative movement through information interaction, so this section will introduce the cooperative topology control package in ``UavNetSim-v1'' platform. The topology control-related package in ``UavNetSim-v1'' is located in the {\ttfamily topology} folder, which in turn contains sub-packages representing different topology control algorithms. Virtual force-based topology control \cite{H.Liu-2010} is supported in the current version. 

Fig. \ref{figure_7} demonstrates the dynamics of the movement processes in ``UavNetSim-v1''. The red solid dots represent UAV nodes, and the black dotted lines represents the communication links between UAVs\footnote{The two UAVs are within the communication range.}. Fig. \ref{figure_7}(a) shows their initial positions, it can be seen that the connectivity of the network is weak in the initial state. After executing the topology control algorithm for a period of time, the connectivity of the network has been greatly improved.

\subsection{Energy Consumption Module}
As described in \cite{Y.Zeng-2019}, the energy consumption of a UAV typically consists of two primary elements: propulsion energy consumption and communication related energy consumption. For the calculation of propulsion energy consumption which is needed to keep the UAV aloft and support its movement, in the current version of ``UavNetSim-v1'', we assume that all UAVs are flying at a constant speed $v$, hence the propulsion power consumption of UAV can be expressed as \cite{Y.Zeng-2019}:
\begin{align}
	P_{{\rm prop}}(v)&=P_0\left(1+\frac{3v^2}{U_{tip}^2}\right)+P_i\left(\sqrt{1+\frac{v^4}{4v_0^4}}-\frac{v^2}{2v_0^2}\right)^{\frac{1}{2}} \nonumber \\
	&+\frac{1}{2}d_0\rho sAv^3,
\end{align}
where $P_0$ and $P_i$ are two constants which denote the blade profile power and induced power in hovering status respectively. The tip speed of the rotor blade is represented by $U_{tip}$. $v_0$ is the mean rotor induced velocity in hover. The fuselage drag ratio and rotor solidity are symbolized by $d_0$ and $s$, respectively. The air density and rotor disc area are represented by $\rho$ and $A$. Therefore, given the movement time $t_{{\rm move}}$, the propulsion energy consumption is calculated as:
\begin{equation}
	E_{{\rm prop}}=P_{{\rm prop}}(v)t_{{\rm move}}.
\end{equation}

For the communication related energy consumption, the current platform considers only the signal radiation energy consumption. Denote the transmit power of UAV as $P_t$, and the transmission time of a packet as $t_{{\rm packet}}$, the communication related energy consumption can be calculated as:
\begin{equation}
	E_{{\rm comm}}=P_tt_{{\rm packet}}.
\end{equation}

\section{Simulator Extensions}
The current version of ``UavNetSim-v1'' can model various components of the UAV networks, including routing/MAC protocols, motion control algorithms, wireless channel and mobility/energy models. Moreover, it can support comprehensive performance evaluation and provide interactive visual analysis. However, several improvements are possible for future research, like:
\begin{itemize}
	\item \textbf{More node types}: In the current version of ``UavNetSim-v1'', we have only implemented UAV entities and their corresponding behaviors. Future work will extend the simulator to support diverse network entities (e.g., satellites, vehicles, sensors). Supporting additional node types will enable broader application scenarios, such as space-air-ground integrated networks, UAV-assisted vehicular networks, and UAV-aided ground data collection.
	\item \textbf{More realistic traffic pattern}: In ``UavNetSim-v1'', traffic is built by a simple random traffic generator function, and the arrival of data packets follows a uniform or Poisson distribution in the current version. Future extensions may include deriving realistic traffic patterns from traces or implementing application and transport layer protocols.
	\item \textbf{Consider security issues}: The security issues of UAV networks have gain considerable attention recently. However, the attack and interference from external malicious nodes are not taken into account in the current version of ``UavNetSim-v1''. A variety of network attacks (e.g., spoofing, jamming, blackhole/wormhole/sybil attacks) and the corresponding secure routing protocols can be considered in the subsequent extensions.
	\item \textbf{More accurate energy consumption model}: The current simulation platform can be extended to include the additional/fewer propulsion caused by UAV acceleration/deceleration. For the calculation of communication related energy consumption, communication circuitry, signal
	processing and signal reception can be taken into account. Furthermore, more energy models \cite{H.V.Abeywickrama-2018} can be integrated into our platform to support different application scenarios.
\end{itemize}

\section{Conclusions}
Due to the prohibitive costs and limited availability of certain network devices, real-world experimentation for UAV networks is usually infeasible, necessitating the use of simulation tools. However, for rapid prototyping, presentations and educational purposes, setting up one of the existing complex network simulators imposes a substantial time and effort burden. Against this background, in this paper, we present ``UavNetSim-v1'', a Python-based open source simulation platform for UAV communication networks.  

``UavNetSim-v1'' provides most of the functionalities developers may need, such as routing protocol modules, MAC protocol modules, and topology control modules, without becoming hard to use. Additionally, this platform supports comprehensive network performance evaluation and offers an interactive visualization interface for analyzing algorithm behavior. To summarize, ``UavNetSim-v1'' can be served as a simple yet powerful alternative to those full-grown network simulators for studying UAV communication networks.

\section{ACKNOWLEDGMENT}
This work was supported in part by the National Key Research and Development Program of China under Grant 2024YFE0107900; in part by the National Natural Science Foundation of China under Grant 62222105; in part by the Natural Science Foundation of Guangdong Province under Grant 2024A1515010235; in part by the Applied Basic Research Funds of Guangzhou under Grant 2024A04J5146; and in part by the Special Program of Leading Entrepreneurship Talent (Team) of ``Yongjiang River'' in Nanning City under Grant 2020003; in part by the Key Technologies for Security Defense and Intelligent Operation and Maintenance of Self healing Distribution communication Network of Power Dispatching and Control Center of Guangdong Power Grid Co. Ltd. under Grant 036000KC23090008(GDKJXM20231043)

\end{document}